\documentclass[prd,twocolumn,showpacs,floatfix,amsmath,amssymb,floatfix,nofootinbib]{revtex4}
\usepackage{graphicx,color,dcolumn,booktabs,bm}
\usepackage{longtable,lscape,comment,braket}
\usepackage{txfonts}
\usepackage{overpic}
\usepackage{amssymb}
\usepackage{indentfirst}
\usepackage{feynmf}   
\usepackage{slashed}  
\usepackage{cases}
\usepackage{epstopdf}
\usepackage{psfrag}
\usepackage{subfigure}
\usepackage{threeparttable}
\pdfoutput=1
\usepackage{color}
\usepackage{multirow}

\def\lrpartial{\buildrel\leftrightarrow\over\partial}
\graphicspath{{Figures/}} %
\usepackage[colorlinks, citecolor=blue,anchorcolor=red,menucolor=red, linkcolor=red,filecolor=red,runcolor=red,urlcolor=blue,frenchlinks=red]{hyperref}

\usepackage{ulem}
\def\be{\begin{equation}}
\def\ee{\end{equation}}

\begin{document}

\title{Potential observation of the $\Upsilon(6S) \to \Upsilon(1^3D_J) \eta$ transitions at Belle II}
\author{Qi Huang$^{1,2}$}
\author{Hao Xu$^{1,2}$}
\author{Xiang Liu$^{1,2}$}\email{xiangliu@lzu.edu.cn}
\author{Takayuki Matsuki$^{3,4}$}\email{matsuki@tokyo-kasei.ac.jp}
\affiliation{$^1$School of Physical Science and Technology, Lanzhou University, Lanzhou 730000, China\\
$^2$Research Center for Hadron and CSR Physics, Lanzhou University and Institute of Modern Physics of CAS, Lanzhou 730000, China\\
$^3$Tokyo Kasei University, 1-18-1 Kaga, Itabashi, Tokyo 173-8602, Japan\\
$^4$Theoretical Research Division, Nishina Center, RIKEN, Wako, Saitama 351-0198, Japan}

\begin{abstract}

We perform the investigation of two-body hidden-bottom transitions of the $\Upsilon(6S)$, which include $\Upsilon(6S) \to \Upsilon(1^3D_J) \eta~~(J=1,2,3)$ decays. For estimating the branching ratios of these processes, we consider contributions from the triangle hadronic loops composed of S-wave $B_{(s)}$ and $B_{(s)}^*$ mesons, which are a bridge to connect the $\Upsilon(6S)$ and final states.
Our results show that both of the branching ratios of these decays can reach $10^{-3}$. Due to such considerable potential to observe these two-body hidden-bottom transitions of the $\Upsilon(6S)$, we suggest the forthcoming Belle II experiment to explore them.

\end{abstract}


\maketitle

\section{Introduction}\label{sec1}

As an updated accelerator with luminosity $8\times 10^{35}$ cm$^{-2}$s$^{-1}$, SuperKEKB
 is currently being constructed. The forthcoming Belle II experiment will accumulate 50 times more data than the previous Belle experiment. It is a good time to explore the potential physical issues close to the Belle II.

Since 2007, experimental studies by the Belle Collaboration have focused on the hadronic transitions of the $\Upsilon(10860)$. The Belle measurement shows that the observed hidden-bottom transitions of $\Upsilon(10860)$ have large branching ratios. For example, all the decay widths of the $\Upsilon(10860) \to \Upsilon(mS) \pi^+ \pi^-~(m\leq3)$ are around $10^{-1}$ MeV and that of the $\Upsilon(10860) \to \Upsilon(1S) \pi^+ \pi^-$ is at least two orders of magnitude larger than those of $\Upsilon(nS) \to \Upsilon(1S) \pi^+ \pi^-~(n\leq4)$ \cite{Abe:2007tk}. In addition, the experimental branching ratios of $\Upsilon(5S) \to \chi_{bJ} \omega~(J=0,1,2)$ transitions can reach up to $10^{-3}$ \cite{He:2014sqj}.
Exploring hidden-bottom dipion decays of the $\Upsilon(10860)$, Belle also discovered two charged bottomonium-like structures $Z_b(10610)$ and $Z_b(10650)$ \cite{Belle:2011aa}.
As indicated in a serial of theoretical studies \cite{Chen:2011zv,Meng:2007tk,Meng:2008dd,Simonov:2008qy,Chen:2011qx,Chen:2014ccr,Chen:2011pv}, the hadronic loop mechanism, which is an equivalent description of coupled-channel effects, may play a crucial role to understand these novel phenomena since the $\Upsilon(10860)$ lies above the $B_{(s)}^{(*)}\bar{B}_{(s)}^{(*)}$ thresholds \cite{Olive:2016xmw}.

Very recently the Belle Collaboration observed $e^+e^-\to \Upsilon_J(1D) \eta$ process and obtained the branching ratio $B(\Upsilon(5S)\to  \Upsilon_J(1D) \eta)=(4.82\pm0.92\pm0.67)\times 10^{-3}$ \cite{Tamponi:2018cuf}, which confirms the predication given in Ref. \cite{Wang:2016qmz}, where the hadronic loop effect was considered in the calculation. This fact again show that the hadronic loop mechanism has important contribution to the hadronic decays of the $\Upsilon(5S)$.

In Particle Data Group (PDG) \cite{Olive:2016xmw}, there is the $\Upsilon(11020)$ above the
$\Upsilon(10860)$. Usually, the $\Upsilon(11020)$ is treated as the $n^{2S+1}L_J=6^3S_1$ state.
Thus, in the following discussions, the $\Upsilon(11020)$ is abbreviated as the $\Upsilon(6S)$ for convenience. Here, we want to emphasize that the $\Upsilon(6S)$ has situation similar to that of the $\Upsilon(10860)$ since  the $\Upsilon(6S)$ is also above the $B_{(s)}^{(*)}\bar{B}_{(s)}^{(*)}$ thresholds. This fact gives us a reason to believe that the coupled-channel effect cannot be ignored when carrying out the studies around the $\Upsilon(6S)$.

When checking the experimental information of the $\Upsilon(6S)$, only the resonance parameters of the $\Upsilon(6S)$ and partial width of $\Upsilon(6S)\to e^+e^-$ are listed, which suggests that experimental study of $\Upsilon(6S)$ is necessary, especially with the running of Belle II. Considering the present status of the $\Upsilon(6S)$, in this work we propose that we firstly investigate two-body hidden-bottom transitions of $\Upsilon(6S)$, i.e., the $\Upsilon(6S)\to \Upsilon(1^3D_J) \eta$ ($J=1,2,3$), which is also motived by the recent observation of $\Upsilon(5S)\to  \Upsilon_J(1D) \eta$ by Belle \cite{Tamponi:2018cuf}.

In this work, using the hadronic loop mechanism, we calculate the branching ratios of the discussed $\Upsilon(6S)\to\Upsilon(1^3D_J) \eta$ decays, by which the potential observation of these two-body hidden-bottom transitions at Belle II can be suggested. We want to emphasize that the present study must become an important part of the physics around the $\Upsilon(6S)$, and further push experimental exploration of these decays.

This paper is organized as follows. After introduction, we present the detailed deductions of $\Upsilon(6S)\to\Upsilon(1^3D_J)\eta$ via the hadronic loop mechanism in Sec. \ref{sec2}. Then, various parameters are determined in Sec.~\ref{sec3}. In Sec. \ref{sec4}, numerical results are presented. Finally the paper will end with a short summary.

\section{The $\Upsilon(6S)\to\Upsilon(1^3D_J)\eta$ transitions via the Hadronic Loop Mechanism}\label{sec2}

\begin{center}
	\begin{figure*}[htbp]
		\scalebox{0.5}{\includegraphics{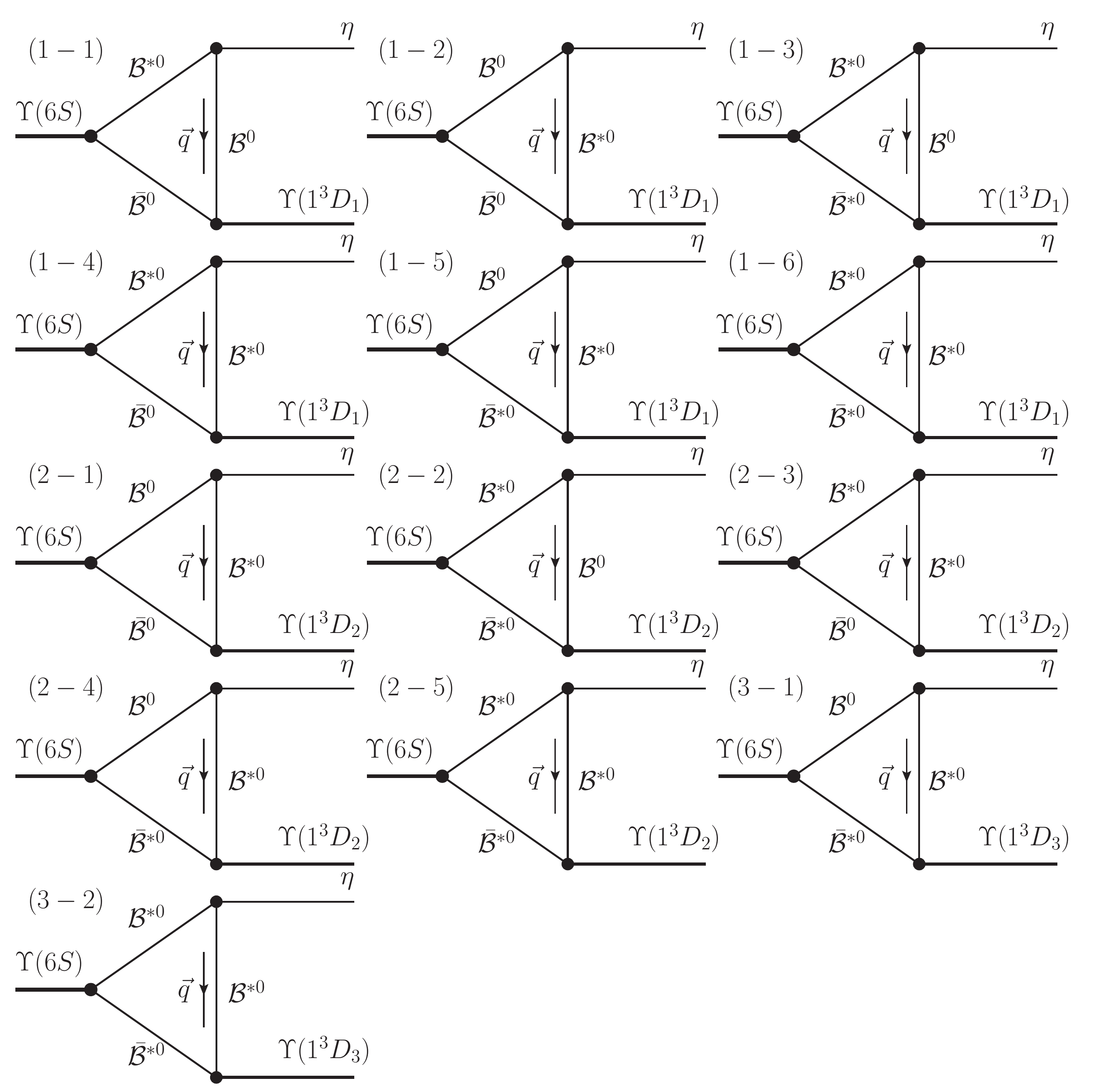}}
		\caption{Schematic diagrams depicting the ${\Upsilon(6S) \to \Upsilon(1^3D_J) \eta}$ transitions via the hadronic loop mechanism.}
		\label{fig:6S-1D-eta}
	\end{figure*}
\end{center}

In this work, we adpot the hadronic loop mechanism to study the $\Upsilon(6S)\to\Upsilon(1^3D_J)\eta$ transitions.
The hadronic loop mechanism has been widely
applied to investigate hidden-bottom/hidden-charm decays of the bottomonium/charmonium \cite{Liu:2006dq,Meng:2008bq,Wang:2016qmz,Wang:2015xsa}.
Under the hadronic loop mechanism, the $\Upsilon(6S)\to\Upsilon(1^3D_J)\eta$ decays may occur via the intermediate triangle loops
composed of S-wave bottom mesons, where all the diagrams depicting this decay process with triangle loops are
  listed in Fig. \ref{fig:6S-1D-eta}.

To calculate the diagrams shown in Fig. \ref{fig:6S-1D-eta}, we adopt the effective Lagrangian approach.
Thus, we firstly need to introduce the relevant effective Lagrangians describing the involved interactions at hadron level.
Considering the constraints from various symmetries, the involved effective Lagrangians can be constructed. For example,
under
the heavy quark spin symmetry in the heavy quark limit, interactions between an S-wave (D-wave) bottomonium and a
 $\mathcal{B}_{(s)}^{(*)} \bar{\mathcal{B}}_{(s)}^{(*)}$ pair can be written as \cite{Wang:2016qmz,Casalbuoni:1996pg}
\begin{eqnarray}\label{eqs:SDHH}
\mathcal{L}_S&=& i g_1 \mathrm{Tr} \left[S^{(Q\bar{Q})} \bar{H}^{(\bar{Q}q)} \gamma^\mu \lrpartial_\mu \bar{H}^{(Q\bar{q})} \right] +H.c., 
\end{eqnarray}
\begin{eqnarray}\label{eqs:SDHH-1}
\mathcal{L}_D &=& i g_2 \mathrm{Tr} \left[D_{\mu\lambda}^{(Q\bar{Q})} \bar{H}^{(\bar{Q}q)} \gamma^\lambda \lrpartial^\mu \bar{H}^{(Q\bar{q})}\right] +H.c.,
\end{eqnarray}
where ${S^{(Q\bar{Q})}}$ and ${D^{(Q\bar{Q})}}$ denote the S-wave  and D-wave multiplets of the bottomonium, respectively \cite{Casalbuoni:1996pg}, which have the concrete expressions
\begin{eqnarray}
S^{(Q\bar{Q})}&=&\frac{1+\slashed{v}}{2}\Big[\Upsilon^\mu\gamma_\mu-\eta_b\gamma_5\Big]\frac{1-\slashed{v}}{2},\nonumber\\\label{J0}
\end{eqnarray}
\begin{eqnarray}
 D^{(Q\bar{Q})\mu\lambda}
&=&\frac{1+\slashed{v}}{2}\Bigg[\Upsilon_3^{\mu\lambda\alpha}\gamma_\alpha+\frac{1}{\sqrt{6}}\Big(\epsilon^{\mu\alpha\beta\rho}v_\alpha\gamma_\beta \Upsilon_{2\rho}^\lambda+\epsilon^{\lambda\alpha\beta\rho}v_\alpha\gamma_\beta \nonumber\\
&& \times\Upsilon_{2\rho}^\mu\Big)
+\frac{\sqrt{15}}{10}\Big[(\gamma^\mu-v^\mu)\Upsilon_1^\lambda+(\gamma^\lambda-v^\lambda)\Upsilon_1^\mu\Big]\nonumber\\
&&-\frac{1}{\sqrt{15}}\Big(g^{\mu\lambda}-v^\mu v^\lambda\Big)\gamma_\alpha\Upsilon_1^\alpha+\eta_{b2}^{\mu\lambda}\gamma_5\Bigg]\frac{1-\slashed{v}}{2}, \label{J2}
\end{eqnarray}
where $v^\mu$ is 4-velocity. $\Upsilon^\mu$ and $\eta_b$ correspond to $S$-wave bottomonia with
$J^{PC}=1^{--}$ and $0^{-+}$, respectively.  $\Upsilon_3$, $\Upsilon_2$, $\Upsilon_1$, and $\eta_{b2}$ denote to $D$-wave
 states in the bottomonium family, which have $J^{PC}=3^{--}$, $2^{--}$, $1^{--}$, and $2^{-+}$, respectively.
For a heavy meson emitting a light Nambu-Goldstone boson, the effective Lagrangian, which is constrained by the heavy quark
symmetry and the chiral symmetry, has the form \cite{Casalbuoni:1996pg,Cheng:1992xi,Yan:1992gz,Wise:1992hn,Burdman:1992gh,Falk}
\begin{eqnarray}\label{eqs:HHP}
\mathcal{L}_{\mathcal{P}} = i g_\pi \mathrm{Tr} \left[ H_b^{(Q\bar{q})} \gamma_\mu \gamma_5 \mathcal{A}_{ba}^\mu \bar{H}_a^{(Q\bar{q})} \right],
\end{eqnarray}
where $\mathcal{A}_\mu = 1/2 (\xi^\dag \partial_\mu \xi - \xi \partial_\mu \xi^\dag)$ with
$\xi = e^{i\mathcal{M_P}/f_\pi}$, and the pseudoscalar octet $\mathcal{M_P}$ reads as
\begin{eqnarray}
\mathcal{M_P} &=&
 \left(
 \begin{array}{ccc}
\sqrt{\frac{1}{2}} \pi^{0}+ \sqrt{\frac{1}{6}} \eta_8 & \pi^{+} & K^{+}\\
\pi^{-} & -\sqrt{\frac{1}{2}} \pi^{0}+ \sqrt{\frac{1}{6}} \eta_8 &  K^{0}\\
 K^{-} & \bar{K}^{0} & -\sqrt{\frac{2}{3}} \eta_8
 \end{array}
 \right).
\end{eqnarray}
In Eqs.~(\ref{eqs:SDHH}), ~(\ref{eqs:SDHH-1}), and (\ref{eqs:HHP}), ${H^{(Q\bar{q})}}$ represents the bottom meson spin doublet ($\mathcal{B}$, $\mathcal{B}^*$)
 \cite{Casalbuoni:1996pg,Kaymakcalan:1983qq,Oh:2000qr,Colangelo:2002mj}, i.e.,
\begin{eqnarray}
H^{(Q\bar{q})}=\frac{1+ \slashed{v}}{2} \left[\mathcal{B}^\ast_\mu \gamma^\mu-\mathcal{B} \gamma^5\right].
\end{eqnarray}
$H^{(\bar{Q}q)}$ corresponds to anti-meson counterpart of ${H^{(Q\bar{q})}}$, which can be obtained by performing the
 charge conjugation transformation.

Further expanding the Lagrangians shown in Eqs. (\ref{eqs:SDHH}) and (\ref{eqs:HHP}), we get the explicit forms of interaction Lagrangians,
i.e.,
\begin{eqnarray}
&&\mathcal{L}_{\Upsilon \mathcal{B}^{(\ast)} \mathcal{B}^{(\ast)}}\nonumber\\
&&= -ig_{\Upsilon \mathcal{BB} } \Upsilon_\mu (\partial^\mu
\mathcal{B} \mathcal{B}^\dagger- \mathcal{B}
\partial^\mu \mathcal{B}^\dagger) \nonumber\\ && \quad+ g_{\Upsilon
\mathcal{B}^\ast \mathcal{B}} \varepsilon^{\mu \nu \alpha \beta}
\partial_\mu \Upsilon_\nu (\mathcal{B}^\ast_\alpha \lrpartial_\beta
\mathcal{B}^\dagger -\mathcal{B} \lrpartial_\beta
\mathcal{B}_\alpha^{\ast \dagger} ) \nonumber\\ && \quad+ ig_{\Upsilon
\mathcal{B}^\ast \mathcal{B}^\ast} \Upsilon^\mu
(\mathcal{B}^\ast_\nu \partial^\nu \mathcal{B}^{\ast \dagger}_\mu
-\partial^\nu \mathcal{B}^{\ast}_\mu \mathcal{B}^{\ast \dagger}_\nu
-\mathcal{B}^\ast_\nu \lrpartial_\mu \mathcal{B}^{\ast \nu
\dagger}), \label{UpsilonBB}
\end{eqnarray}
\begin{eqnarray}
&&\mathcal{L}_{\Upsilon_J \mathcal{B}^{(\ast)}\mathcal{B}^{(\ast)}}\nonumber\\
&&=g_{\Upsilon_1 \mathcal{B}\mathcal{B}}\Upsilon_1^\mu(\mathcal{B}^\dag\partial_\mu \mathcal{B}-\mathcal{B}\partial_\mu \mathcal{B}^\dag)\nonumber\\
&&\quad+ig_{\Upsilon_1 \mathcal{B}\mathcal{B}^\ast}\epsilon^{\mu\nu\alpha\beta}\Big[\mathcal{B}^\dag\overleftrightarrow{\partial}_\mu \mathcal{B}_\beta^\ast-\mathcal{B}_\beta^{\ast\dag}\overleftrightarrow{\partial}_\mu \mathcal{B}\Big]\partial_\nu\Upsilon_{1\alpha}\nonumber\\
&&\quad+g_{\Upsilon_1 \mathcal{B}^\ast \mathcal{B}^\ast}\Big[-4(\Upsilon_1^\mu \mathcal{B}^{\ast\nu}\partial_\mu \mathcal{B}_\nu^{\ast\dag}-\Upsilon_1^\mu \mathcal{B}_\nu^{\ast\dag}\partial_\mu \mathcal{B}^{\ast\nu})\nonumber\\
&&\quad+\Upsilon_1^\mu \mathcal{B}^{\ast\nu}\partial_\nu \mathcal{B}^{\ast\dag}_\mu-\Upsilon_1^\mu \mathcal{B}^{\ast\nu\dag}\partial_\nu \mathcal{B}^{\ast}_\mu\Big]\nonumber\\
&&\quad+ig_{\Upsilon_{2} \mathcal{B}\mathcal{B}^\ast}\Upsilon_{2}^{\mu\nu}(\mathcal{B}^\dag\overleftrightarrow{\partial}_\nu \mathcal{B}_\mu^\ast-\mathcal{B}_\mu^{\ast\dag}\overleftrightarrow{\partial}_\nu \mathcal{B})\nonumber\\
&&\quad+g_{\Upsilon_{2} \mathcal{B}^\ast \mathcal{B}^\ast}\epsilon_{\alpha\beta\mu\nu}\Big[\mathcal{B}^{\ast^\nu\dag}\overleftrightarrow{\partial}^\beta \mathcal{B}_\lambda^{\ast}+\mathcal{B}^{\ast^\nu}\overleftrightarrow{\partial}^\beta \mathcal{B}_\lambda^{\ast\dag}\Big]\partial^\mu\Upsilon_{2}^{\alpha\lambda}\nonumber\\
&&\quad+g_{\Upsilon_3\mathcal{B}^\ast \mathcal{B}^\ast}\Upsilon_3^{\mu\nu\alpha}\Big[\mathcal{B}_\alpha^{\ast\dag}\overleftrightarrow{\partial}_\mu \mathcal{B}_\nu^\ast+\mathcal{B}_\nu^{\ast\dag}\overleftrightarrow{\partial}_\mu \mathcal{B}_\alpha^\ast\Big],\\
&&\mathcal{L}_{\mathcal{B}^{(\ast)}\mathcal{B}^{(\ast)} \eta_8}\nonumber\\
&&= i g_{\mathcal{B} \mathcal{B}^* \eta_8} (\mathcal{B}^\dag \partial_\mu \eta_8 \mathcal{B}^{*\mu} - \mathcal{B}^{*\dag\mu} \partial_\mu \eta_8 \mathcal{B})\nonumber\\
&&\quad - g_{\mathcal{B}^* \mathcal{B}^* \eta_8} \varepsilon^{\mu\nu\alpha\beta} \partial_\mu \mathcal{B}^{*\dag}_\nu \partial_\alpha \mathcal{B}^*_\beta \eta_8,
\end{eqnarray}
where $\mathcal{B}^{(\ast)\dag}$ and $\mathcal{B}^{(\ast)}$ are defined as $\mathcal{B}^{(\ast)\dag}=(B^{(\ast)+},B^{(\ast)0},B_s^{(\ast)0})$ and
 $\mathcal{B}^{(\ast)}=(B^{(\ast)-},\bar{B}^{(\ast)0},\bar{B}_s^{(\ast)0})^T$, respectively.

Using the Lagrangians above,  we can write out the amplitudes of the processes $\Upsilon(6S) \to \Upsilon(1^3D_J) \eta$ with $J=1,2,3$, which are depicted in Fig. \ref{fig:6S-1D-eta}.
As for the $\Upsilon(6S) \to \Upsilon(1^3D_1)\eta$ transition, with $\tilde{g}^{\mu\nu}(p,m_p) \equiv -g^{\mu\nu}+\frac{p^\mu p^\nu}{m_p^2}$,
 the amplitudes are,
\begin{eqnarray}
\mathcal{M}_{(1-1)} &=& \int \frac{d^4q}{(2\pi)^4} \left[g_{\Upsilon  \mathcal{B}^* \bar{\mathcal{B}}} \varepsilon^{\mu\nu\alpha\beta} p_{1\mu} \epsilon_{\Upsilon \nu} (k_{1\beta} \right. \nonumber\\
&& \left. -k_{2\beta})\right] \left[g_{\mathcal{B}^* \bar{\mathcal{B}} \eta} p_{2\lambda}\right] \left[ i g_{\bar{\mathcal{B}} \mathcal{B} \Upsilon_1 } \epsilon^{*}_{\Upsilon_1 \zeta } (k_{2}^\zeta-q^\zeta) \right]\nonumber\\
&&\times\frac{\tilde{g}_\alpha^\lambda(k_1,m_{\mathcal{B}^*})}{k_1^2-m_{\mathcal{B}^*}^2} \frac{1}{k_2^2-m_{\mathcal{B}}^2} \frac{1}{q^2-m_{\mathcal{B}}^2}\mathcal{F}^2(q^2),
\end{eqnarray}

\begin{eqnarray}
\mathcal{M}_{(1-2)} &=& \int \frac{d^4q}{(2\pi)^4} \left[g_{\Upsilon  \mathcal{B} \bar{\mathcal{B}}} \epsilon_{\Upsilon \mu} (k_2^\mu-k_1^\mu)\right] \left[g_{\mathcal{B} \bar{\mathcal{B}}^* \eta} p_{2\lambda}\right]\nonumber\\
&&\times \left[i g_{\bar{\mathcal{B}} \mathcal{B}^* \Upsilon_1 } \varepsilon^{\zeta\eta\kappa\xi} p_{3\eta} \epsilon^{*}_{\Upsilon_1 \kappa} (k_{2\zeta} -q_\zeta)\right]\nonumber\\
&&\times\frac{1}{k_1^2-m_{\mathcal{B}}^2} \frac{1}{k_2^2-m_{\mathcal{B}}^2} \frac{\tilde{g}_\xi^\lambda(q,m_{\mathcal{B}^*})}{q^2-m_{\mathcal{B}^*}^2}\mathcal{F}^2(q^2),
\end{eqnarray}

\begin{eqnarray}
\mathcal{M}_{(1-3)} &=& \int \frac{d^4q}{(2\pi)^4} \left[g_{\Upsilon  \mathcal{B}^* \bar{\mathcal{B}}^*} \epsilon_{\Upsilon }^\mu \left(g_{\mu\alpha}k_{2\beta}- g_{\mu\beta}k_{1\alpha}\right.\right.\nonumber\\
&&\left.\left. + g_{\alpha\beta} (k_{1\mu}-k_{2\mu})\right)\right] \left[g_{\mathcal{B}^* \bar{\mathcal{B}} \eta}p_{2\lambda}\right]\nonumber\\
&&\times \left[i g_{\bar{\mathcal{B}}^* \mathcal{B} \Upsilon_1 } \varepsilon^{\zeta\eta\kappa\xi} p_{3\eta} \epsilon^{*}_{\Upsilon_1 \kappa} (k_{2\zeta}-q_\zeta)\right]\nonumber\\
&&\times\frac{\tilde{g}^{\beta\lambda}(k_1,m_{\mathcal{B}^*})}{k_1^2-m_{\mathcal{B}^*}^2} \frac{\tilde{g}_\xi^\alpha(k_2,m_{\mathcal{B}^*})}{k_2^2-m_{\mathcal{B}^*}^2} \frac{1}{q^2-m_{\mathcal{B}}^2} \mathcal{F}^2(q^2),
\end{eqnarray}

\begin{eqnarray}
\mathcal{M}_{(1-4)} &=& \int \frac{d^4q}{(2\pi)^4} \left[g_{\Upsilon  \mathcal{B}^* \bar{\mathcal{B}}} \varepsilon^{\mu\nu\alpha\beta} p_{1\mu} \epsilon_{\Upsilon \nu} (k_{2\beta}-k_{1\beta})\right]\nonumber\\
&&\times\left[g_{\mathcal{B}^* \bar{\mathcal{B}}^* \eta} \varepsilon^{\lambda\rho\delta\sigma} k_{1\lambda} q_\delta\right]\left[i g_{\bar{\mathcal{B}} \mathcal{B}^* \Upsilon_1 }\varepsilon^{\zeta\eta\kappa\xi} p_{3\eta}\right.\nonumber\\
&&\left.\times \epsilon^{*}_{\Upsilon_1 \kappa} (k_{2\zeta}-q_\zeta)\right]\frac{\tilde{g}_{\alpha\rho}(k_1,m_{\mathcal{B}^*})}{k_1^2-m_{\mathcal{B}^*}^2} \frac{1}{k_2^2-m_{\mathcal{B}}^2}\nonumber\\
&&\times \frac{\tilde{g}_{\sigma\xi}(q,m_{\mathcal{B}^*})}{q^2-m_{\mathcal{B}^*}^2}\mathcal{F}^2(q^2),
\end{eqnarray}

\begin{eqnarray}
\mathcal{M}_{(1-5)} &=& \int \frac{d^4q}{(2\pi)^4} \left[g_{\Upsilon  \mathcal{B} \bar{\mathcal{B}}^*} \varepsilon^{\mu\nu\alpha\beta} p_{1\mu} \epsilon_{\Upsilon \nu} (k_{1\beta} \right.\nonumber\\
&&\left.-k_{2\beta})\right]\left[g_{\mathcal{B} \bar{\mathcal{B}}^* \eta} p_{2\lambda}\right] \left[i g_{\bar{\mathcal{B}}^* \mathcal{B}^* \Upsilon_1 } \epsilon^{*}_{\Upsilon_1  \zeta} \left(4 g_{\kappa\xi} \right. \right.\nonumber\\
&&\times \left. \left.(k_{2}^\zeta -q^\zeta) + g^{\zeta}_\xi q_\kappa - g^{\zeta}_\kappa k_{2\xi}\right)\right]\nonumber\\
&&\times\frac{1}{k_1^2-m_{\mathcal{B}}^2} \frac{\tilde{g}_\alpha^\kappa(k_2,m_{\mathcal{B}^*})}{k_2^2-m_{\mathcal{B}^*}^2} \frac{\tilde{g}^{\lambda\xi}(q,m_{\mathcal{B}^*})}{q^2-m_{\mathcal{B}^*}^2} \mathcal{F}^2(q^2),
\end{eqnarray}

\begin{eqnarray}
\mathcal{M}_{(1-6)} &=& \int \frac{d^4q}{(2\pi)^4} \left[g_{\Upsilon  \mathcal{B}^* \bar{\mathcal{B}}^*} \epsilon_{\Upsilon }^\mu \left(g_{\mu\alpha}k_{2\beta}- g_{\mu\beta}k_{1\alpha} \right.\right.\nonumber\\
&&\left.\left.+ g_{\alpha\beta} (k_{1\mu}-k_{2\mu})\right)\right] \left[g_{\mathcal{B}^* \bar{\mathcal{B}}^* \eta} \varepsilon^{\lambda\rho\delta\sigma} k_{1\lambda} q_\delta\right]\nonumber\\
&&\times\left[i g_{\bar{\mathcal{B}}^* \mathcal{B}^* \Upsilon_1 } \epsilon^{*\zeta}_{\Upsilon_1 } \left(4 g_{\kappa\xi} (k_{2\zeta}-q_\zeta) + g_{\zeta\xi} q_\kappa\right.\right.\nonumber\\
&&\left.\left. - g_{\zeta\kappa} k_{2\xi}\right)\right] \frac{\tilde{g}_\rho^\beta(k_1,m_{\mathcal{B}^*})}{k_1^2-m_{\mathcal{B}^*}^2} \frac{\tilde{g}^{\alpha\kappa}(k_2,m_{\mathcal{B}^*})}{k_2^2-m_{\mathcal{B}^*}^2}\nonumber\\
&&\times \frac{\tilde{g}_\sigma^\xi(q,m_{\mathcal{B}^*})}{q^2-m_{\mathcal{B}^*}^2} \mathcal{F}^2(q^2).
\end{eqnarray}

As for the $\Upsilon(6S) \to \Upsilon(1^3D_2)\eta$ transition, the amplitudes read as,
\begin{eqnarray}
\mathcal{M}_{(2-1)} &=& \int \frac{d^4q}{(2\pi)^4} \left[g_{\Upsilon  \mathcal{B} \bar{\mathcal{B}}} \epsilon_{\Upsilon \mu} (k_2^\mu-k_1^\mu)\right]\left[g_{\mathcal{B} \bar{\mathcal{B}}^* \eta}   p_{2\lambda} \right]\nonumber\\
&&\times \left[g_{\bar{\mathcal{B}} \mathcal{B}^* \Upsilon_2 } \epsilon^{*}_{\Upsilon_2 \zeta\eta} (k_{2}^\eta-q^\eta)\right] \frac{1}{k_1^2-m_{\mathcal{B}}^2}\nonumber\\
&&\times \frac{1}{k_2^2-m_{\mathcal{B}}^2} \frac{\tilde{g}^{\zeta\lambda}(q,m_{\mathcal{B}^*})}{q^2-m_{\mathcal{B}^*}^2}\mathcal{F}^2(q^2),
\end{eqnarray}

\begin{eqnarray}
\mathcal{M}_{(2-2)} &=& \int \frac{d^4q}{(2\pi)^4} \left[g_{\Upsilon  \mathcal{B}^* \bar{\mathcal{B}}^*} \epsilon_{\Upsilon }^\mu \left(g_{\mu\alpha}k_{2\beta}- g_{\mu\beta}k_{1\alpha} \right.\right.\nonumber\\
&&\left.\left. + g_{\alpha\beta} (k_{1\mu}-k_{2\mu})\right)\right] \left[g_{\mathcal{B}^* \bar{\mathcal{B}} \eta} p_{2\lambda}\right] \left[g_{\bar{\mathcal{B}}^* \mathcal{B} \Upsilon_2 } \right.\nonumber\\
&&\left.\times\epsilon^{*}_{\Upsilon_{2\zeta\eta}} (k_{2}^\eta-q^\eta)\right] \frac{\tilde{g}^{\beta\lambda}(k_1,m_{\mathcal{B}^*})}{k_1^2-m_{\mathcal{B}^*}^2}\nonumber\\
&&\times \frac{\tilde{g}^{\zeta\alpha}(k_2,m_{\mathcal{B}^*})}{k_2^2-m_{\mathcal{B}^*}^2} \frac{1}{q^2-m_{\mathcal{B}}^2}\mathcal{F}^2(q^2),
\end{eqnarray}

\begin{eqnarray}
\mathcal{M}_{(2-3)} &=& \int \frac{d^4q}{(2\pi)^4} \left[g_{\Upsilon  \mathcal{B}^* \bar{\mathcal{B}}} \varepsilon^{\mu\nu\alpha\beta} p_{1\mu} \epsilon_{\Upsilon  \nu} (k_{1\beta}-k_{2\beta})\right]\nonumber\\
&& \times \left[g_{\mathcal{B}^* \bar{\mathcal{B}}^* \eta} \varepsilon^{\lambda\rho\delta\sigma} k_{1\lambda} q_\delta\right] \left[g_{\bar{\mathcal{B}} \mathcal{B}^* \Upsilon_2 } \epsilon^{*}_{\Upsilon_2 \zeta\eta} (q^\eta-k_{2}^\eta)\right]\nonumber\\
&&\times \frac{\tilde{g}_{\alpha\rho}(k_1,m_{\mathcal{B}^*})}{k_1^2-m_{\mathcal{B}^*}^2} \frac{1}{k_2^2-m_{\mathcal{B}}^2} \frac{\tilde{g}_{\sigma}^\zeta(k_1,m_{\mathcal{B}^*})}{q^2-m_{\mathcal{B}^*}^2} \mathcal{F}^2(q^2),
\end{eqnarray}

\begin{eqnarray}
\mathcal{M}_{(2-4)} &=& \int \frac{d^4q}{(2\pi)^4} \left[g_{\Upsilon  \mathcal{B} \bar{\mathcal{B}}^*} \varepsilon^{\mu\nu\alpha\beta} p_{1\mu} \epsilon_{\Upsilon  \nu} (k_{1\beta}-k_{2\beta})\right]\nonumber\\
&& \times\left[g_{\mathcal{B} \bar{\mathcal{B}}^* \eta} p_{2\lambda}\right] \left[g_{\bar{\mathcal{B}}^* \mathcal{B}^* \Upsilon_2 } \varepsilon^{\kappa\xi\zeta\eta} p_{3\zeta} \epsilon^{*}_{\Upsilon_2 \kappa\omega} (g_{\tau}^\omega \right.\nonumber\\
&&\left.\times g_{\chi\eta} - g_{\tau\eta} g_{\chi}^\omega) (q_\xi-k_{2\xi})\right] \frac{1}{k_1^2-m_{\mathcal{B}}^2} \nonumber\\
&&\times\frac{\tilde{g}_{\alpha}^{\tau}(k_2,m_{\mathcal{B}^*})}{k_2^2-m_{\mathcal{B}^*}^2} \frac{\tilde{g}^{\lambda\xi}(q,m_{\mathcal{B}^*})}{q^2-m_{\mathcal{B}^*}^2}\mathcal{F}^2(q^2),
\end{eqnarray}

\begin{eqnarray}
\mathcal{M}_{(2-5)} &=& \int \frac{d^4q}{(2\pi)^4} \left[g_{\Upsilon  \mathcal{B}^* \bar{\mathcal{B}}^*} \epsilon_{\Upsilon }^\mu \left(g_{\mu\alpha}k_{2\beta} - g_{\mu\beta}k_{1\alpha}\right.\right.\nonumber\\
&&\left.\left. + g_{\alpha\beta} (k_{1\mu}-k_{2\mu})\right)\right] \left[g_{\mathcal{B}^* \bar{\mathcal{B}}^* \eta} \varepsilon^{\lambda\rho\delta\sigma} k_{1\lambda} q_\delta\right]\nonumber\\
&&\times\left[g_{\bar{\mathcal{B}}^* \mathcal{B}^* \Upsilon_2 } \varepsilon^{\kappa\xi\zeta\eta} p_{3\zeta} \epsilon^{*}_{\Upsilon_2 \kappa\omega} (g_{\tau}^\omega g_{\chi\eta} \right.\nonumber\\
&&\left. - g_{\tau\eta} g_{\chi}^\omega) (q_\xi-k_{2\xi})\right] \frac{\tilde{g}_{\rho}^{\beta}(k_1,m_{\mathcal{B}^*})}{k_1^2-m_{\mathcal{B}^*}^2} \nonumber\\
&&\times\frac{\tilde{g}^{\alpha\tau}(k_2,m_{\mathcal{B}^*})}{k_2^2-m_{\mathcal{B}^*}^2} \frac{\tilde{g}_{\sigma}^{\xi}(q,m_{\mathcal{B}^*})}{q^2-m_{\mathcal{B}^*}^2} \mathcal{F}^2(q^2).
\end{eqnarray}

As for the $\Upsilon(6S) \to \Upsilon(1^3D_3)\eta$ transition, the amplitudes can be expressed as
\begin{eqnarray}
\mathcal{M}_{(3-1)} &=& \int \frac{d^4q}{(2\pi)^4} \left[g_{\Upsilon  \mathcal{B} \bar{\mathcal{B}}^*} \varepsilon^{\mu\nu\alpha\beta} p_{1\mu} \epsilon_{\Upsilon \nu} (k_{1\beta}-k_{2\beta})\right]\nonumber\\
&&\times\left[g_{\mathcal{B} \bar{\mathcal{B}}^* \eta} p_{2\lambda}\right]\left[i g_{\bar{\mathcal{B}}^* \mathcal{B}^* \Upsilon_3 } \epsilon^{*}_{\Upsilon_3 \zeta\eta\kappa} (g_{\tau}^{\eta} g_{\omega}^{\kappa}\right.\nonumber\\
&&\left.+ g_{\tau}^{\kappa} g_{\omega}^{\eta}) (k_{2}^{\zeta}-q^\zeta)\right] \frac{1}{k_1^2-m_{\mathcal{B}}^2} \frac{\tilde{g}_{\alpha}^{\tau}(k_2,m_{\mathcal{B}^*})}{k_2^2-m_{\mathcal{B}^*}^2}\nonumber\\
&&\times  \frac{\tilde{g}^{\lambda\omega}(q,m_{\mathcal{B}^*})}{q^2-m_{\mathcal{B}^*}^2} \mathcal{F}^2(q^2),
\end{eqnarray}

\begin{eqnarray}
\mathcal{M}_{(3-2)} &=& \int \frac{d^4q}{(2\pi)^4} \left[g_{\Upsilon  \mathcal{B}^* \bar{\mathcal{B}}^*} \epsilon_{\Upsilon }^\mu \left(g_{\mu\alpha}k_{2\beta}- g_{\mu\beta}k_{1\alpha} + g_{\alpha\beta}  \right.\right.\nonumber\\
&&\left.\left.\times (k_{1\mu} -k_{2\mu})\right)\right] \left[g_{\mathcal{B}^* \bar{\mathcal{B}}^* \eta} \varepsilon^{\lambda\rho\delta\sigma} k_{1\lambda} q_\delta\right] \left[i g_{\bar{\mathcal{B}}^* \mathcal{B}^* \Upsilon_3 }\right. \nonumber\\
&&\left.\times \epsilon^{*}_{\Upsilon_3 \zeta\eta\kappa} (g_{\tau}^{\eta}g_{\omega}^{\kappa} + g_{\tau}^{\kappa} g_{\omega}^{\eta}) (k_{2}^{\zeta}-q^\zeta)\right] \frac{\tilde{g}_\rho^\beta(k_1,m_{\mathcal{B}^*})}{k_1^2-m_{\mathcal{B}^*}^2}\nonumber\\
&&\times \frac{\tilde{g}^{\alpha\tau}(k_2,m_{\mathcal{B}^*})}{k_2^2-m_{\mathcal{B}^*}^2} \frac{\tilde{g}_\sigma^\rho(q,m_{\mathcal{B}^*})}{q^2-m_{\mathcal{B}^*}^2}\mathcal{F}^2(q^2).
\end{eqnarray}
In the above amplitudes, the monopole form factor
$\mathcal{F}(q^2) = (m_E^2 - \Lambda^2)/(q^2 - \Lambda^2)$ is introduced in our calculation since
the structure effect of the interaction vertex cannot be ignored and off-shell effect from the
 exchanged bottom mesons in triangle loops should be compensated by this way. This type of form factor is supported by the QCD sum rule study \cite{Gortchakov:1995im}. Here, $m_E$ is the mass of the exchanged bottom meson
$\mathcal{B}^{(*)}$, and the cutoff $\Lambda$ is parameterized as $\Lambda = m_E + \alpha_\Lambda \Lambda_{QCD}$ with
$\Lambda_{QCD}=0.22$ GeV (see Refs.~\cite{Liu:2006dq,Liu:2009dr,Li:2013zcr} for more details).

Finally, the total sum of amplitudes reads as
\begin{eqnarray}\label{TDA}
\mathcal{M}_J^{\mathrm{Total}} = 4 \sum_{j} \mathcal{M}^q_{(J-j)} + 2 \sum_{j} \mathcal{M}^s_{(J-j)}\label{rhs}
\end{eqnarray}
with ${J=1,2,3}$, which correspond to the $\Upsilon(6S) \to \Upsilon(1^3D_1) \eta$,
$\Upsilon(6S) \to \Upsilon(1^3D_2) \eta$, and $\Upsilon(6S) \to \Upsilon(1^3D_3) \eta$ transitions, respectively.
The triangle loops can be composed of either bottom mesons or bottom-strange mesons. To distinguish them, we
adopt the superscripts in amplitudes $\mathcal{M}^q_{(J-j)}$ and $\mathcal{M}^s_{(J-j)}$.
The factor 4 in the first term of right-hand side of Eq. (\ref{rhs}) comes from the charge
conjugation transformation $B^{(\ast)}\leftrightarrow\bar{B}^{(\ast)}$ and the isospin transformations $B^{(\ast)0}\leftrightarrow B^{(\ast)+}$
and $\bar{B}^{(\ast)0}\leftrightarrow B^{(\ast)-}$, while the factor 2 in the second term of right-hand side of Eq. (\ref{rhs}) is due to the charge
conjugation transformation $B_s^{(\ast)}\leftrightarrow \bar{B}_s^{(\ast)}$.

With the total sum of amplitudes for each process, the general expression of the partial decay widths is
\begin{eqnarray}
\Gamma_J = \frac{1}{3} \frac{1}{8\pi} \frac{|\vec{p}_\eta|}{m_{\Upsilon(6S)}^2}
|\overline{\mathcal{M}^{\mathrm{Total}}_J}|^2,
\end{eqnarray}
which averages over the polarization of initial $\Upsilon(6S)$ and sums over the polarizations of the $\Upsilon(1^3D_J)$.

\section{Input parameters}\label{sec3}
Before displaying our results, we have to illustrate various parameters including masses and coupling constants used in this work.
First, we need the input of the masses.
For the masses of the bottomonia $\Upsilon(1^3D_1)$ and $\Upsilon(1^3D_3)$, 10.153 GeV \cite{Eichten:1980mw,Wang:2016qmz}
and 10.174 GeV \cite{Eichten:1980mw,Wang:2016qmz} are adopted, respectively,
whereas PDG values  \cite{Olive:2016xmw}  are used for other bottomonia involved in this work.

Utilizing the partial decay widths of $\Upsilon(6S) \to B_{(s)}^{(*)}\bar{B}_{(s)}^{(*)}$ estimated in Ref.~\cite{Godfrey:2015dia},
we can obtain the coupling constants $g_{\Upsilon(6S) B_{(s)}^{(*)}\bar{B}_{(s)}^{(*)}}$ in Eq.~(\ref{UpsilonBB}).
In Table \ref{tab:cc-6S-HH} we list the calculated partial decay widths in Ref. \cite{Godfrey:2015dia} as
well as the extracted coupling constants \cite{Huang:2017kkg}.
 \renewcommand{\arraystretch}{1.8}
\begin{table}[htpb]
\centering \caption{The partial decay widths given in Ref.~\cite{Godfrey:2015dia} and the extracted coupling constants
	 $g_{\Upsilon(6S) B_{(s)}^{(*)}\bar{B}_{(s)}^{(*)}}$ \cite{Huang:2017kkg}.\label{tab:cc-6S-HH}}
\begin{tabular}{ccccc}
\toprule[1pt]
Final state  &~& Decay width (MeV) &~& Coupling constant\\
\midrule[0.6pt] %
$B \bar{B}$ &~& 1.32 &~& 0.654\\
$B \bar{B}^{*}$ &~& 7.59 &~& $0.077~\mathrm{GeV}^{-1}$\\
$B^* \bar{B}^*$ &~& 5.89 &~& 0.611\\
$B_s \bar{B}_s$ &~& $1.31\times10^{-3}$ &~& 0.043\\
$B_s \bar{B}_s^{*}$ &~& 0.136 &~& $0.023~\mathrm{GeV}^{-1}$\\
$B_s^* \bar{B}_s^*$ &~& 0.310 &~& 0.354\\
\bottomrule[1pt]
\end{tabular}
\end{table}

The coupling constants $g_{\Upsilon(1^3D_J) B_{(s)}^{(*)}\bar{B}_{(s)}^{(*)}}$
are related to one coupling $g_2$ of Eq. (\ref{eqs:SDHH}) under the heavy quark symmetry:
\begin{eqnarray}
g_{\Upsilon_1 \mathcal{B} \mathcal{B}}&=&-2g_2\frac{\sqrt{15}}{3}\sqrt{m_{\Upsilon_1}m_{\mathcal{B}}m_{\mathcal{B}}},\label{hhh1}\nonumber\\
g_{\Upsilon_1 \mathcal{B}\mathcal{B}^\ast}&=&g_2\frac{\sqrt{15}}{3}\sqrt{m_\mathcal{B} m_{\mathcal{B}^\ast}/m_{\Upsilon_1}},\label{hhh3}\nonumber\\
g_{\Upsilon_1 \mathcal{B}^\ast \mathcal{B}^\ast}&=&g_2\frac{\sqrt{15}}{15}\sqrt{m_{\Upsilon_1}m_{\mathcal{B}^\ast}m_{\mathcal{B}^\ast}},\nonumber\\
g_{\Upsilon_{2} \mathcal{B}\mathcal{B}^\ast}&=&2g_2\sqrt{\frac{3}{2}}\sqrt{m_{\Upsilon_{2}}m_\mathcal{B} m_{\mathcal{B}^\ast}},\label{eq15}\nonumber\\
g_{\Upsilon_{2} \mathcal{B}^\ast \mathcal{B}^\ast}&=&-2g_2\sqrt{\frac{1}{6}}\sqrt{m_{\mathcal{B}^\ast} m_{\mathcal{B}^\ast}/m_{\Upsilon_{2}}},\label{eq16}\nonumber\\
g_{\Upsilon_3\mathcal{B}^\ast \mathcal{B}^\ast}&=&2g_2\sqrt{m_{\Upsilon_3}m_{\mathcal{B}^\ast}m_{\mathcal{B}^\ast}}.\label{hhh2}\nonumber
\end{eqnarray}

We now have to determine the last remaining coupling constant $g_2$.
For $g_2$, we fix it as follows \cite{Wang:2016qmz}: we first define the decay constant of the vector meson $\Upsilon(1^3D_1)$ \cite{Neubert:1992}
\begin{eqnarray}\label{ee}
\langle0|Q\gamma^\mu\bar{Q}|V\rangle=f_V M_V \epsilon^\mu_V,
\end{eqnarray}
with $M_V$ and $\epsilon^\mu_V$ being the mass and the polarization vector of $\Upsilon(1^3D_1)$, respectively.
Then, using the relation $g_{VBB}\simeq M_{V}/f_{V}$ under the vector meson dominance ansatz
\cite{Colangelo:2003sa,Lin:1999ad,Deandrea:2003pv}, we can get $g_{\Upsilon(1^3D_1)BB}$, and $g_2$ eventually.

The decay constant $f_V$ can be obtained by fitting to the leptonic decay width $\Gamma[\Upsilon(1^3D_1)\to e^+e^-]=1.38$ eV \cite{Godfrey:2015dia}.
Using the relation \cite{Neubert:1992}
\begin{eqnarray}
\Gamma_{V\to e^+e^-}=\frac{4\pi}{3}\frac{\alpha^2}{M_V}f_V^2C_V,
\end{eqnarray}
where $\alpha$ is the fine-structure constant and $C_V=1/9$ for the $\Upsilon(1^3D_1)$ meson, we get
$f_{V}=23.8$ MeV. Therefore, we get
$g_2=9.83 \textnormal{ GeV}^{-3/2}$.

We now turn to the coupling constants between $\eta$ and $B_{(s)}^{(*)}\bar{B}_{(s)}^{(*)}$. Relations can be extracted from Eq. (\ref{eqs:HHP}):
\begin{eqnarray}
\frac{g_{BB^*\eta_8}}{\sqrt{m_B m_{B^*}}} &=& g_{B^* B^* \eta_8} = \frac{2}{\sqrt{6}} \frac{g_\pi}{f_\pi}\nonumber,\\
\frac{g_{B_sB_s^*\eta_8}}{\sqrt{m_{B_s} m_{B_s^*}}} &=& g_{B_s^* B_s^* \eta_8} = -2 \sqrt{\frac{2}{3}}\frac{g_\pi}{f_\pi}\nonumber,
\end{eqnarray}
with $f_\pi=131$ MeV and $g_\pi=0.569$. Furthermore, $\eta$ is a mixture between octet $\eta_8$ and singlet $\eta_1$:
\begin{eqnarray}
|\eta\rangle=\cos\theta|\eta_8\rangle-\sin\theta|\eta_1\rangle,\ \ \ |\eta^\prime\rangle=\sin\theta|\eta_8\rangle+\cos\theta|\eta_1\rangle,
\end{eqnarray}
where $\theta=-19.1^\circ$ is fixed by experimental data \cite{Coffman:1988ve,Jousset:1988ni}.

\section{Estimate of branching ratios}\label{sec4}
\begin{center}
\begin{figure}[htbp]
\scalebox{0.34}[0.32]{\includegraphics{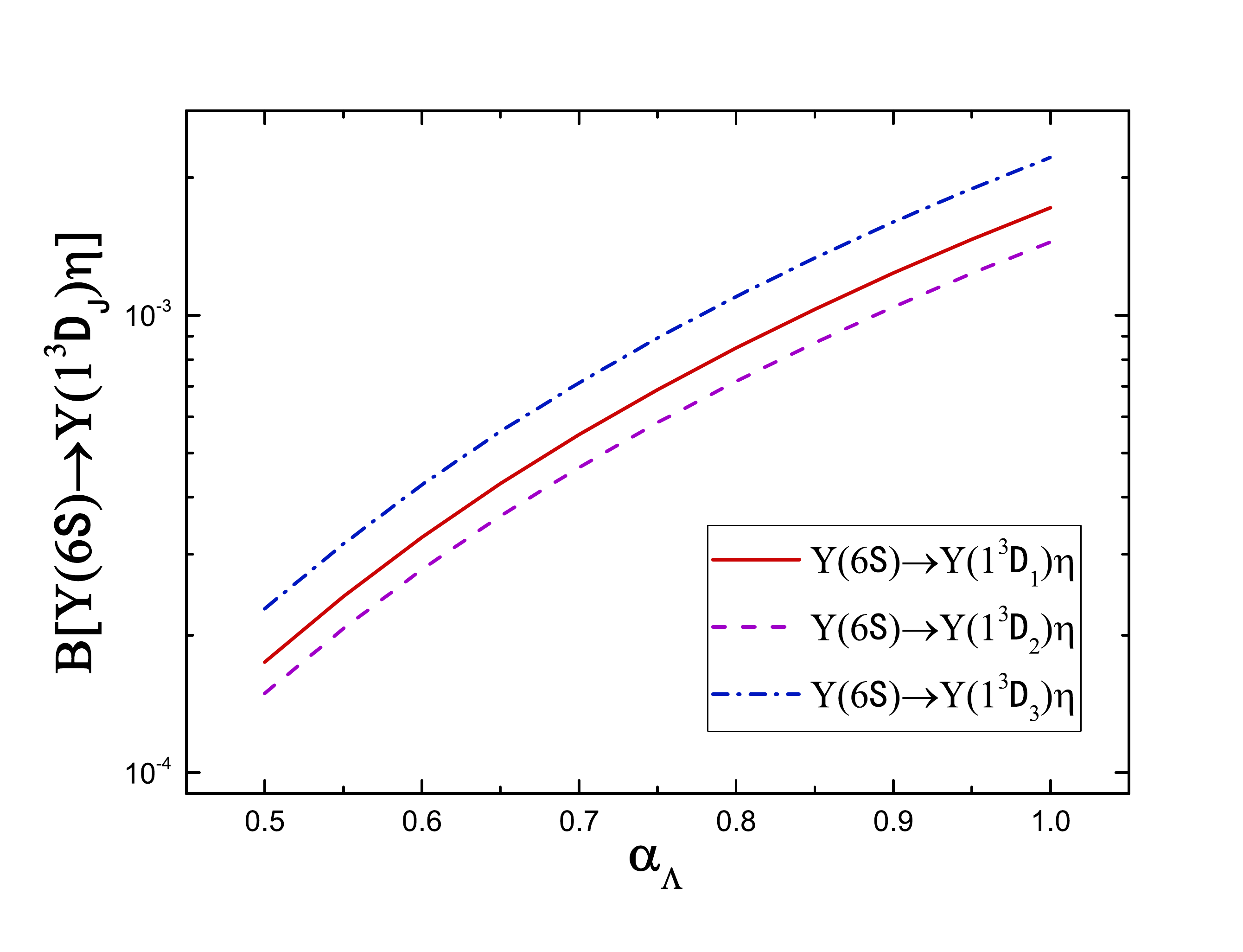}}
\caption{The $\alpha_\Lambda$ dependence of the branching ratios for the processes $\Upsilon(6S) \to \Upsilon(1^3D_J) \eta~(J=1,2,3)$.}
\label{fig:B123-13DJ}
\end{figure}
\end{center}

\begin{center}
\begin{figure}[htbp]
\scalebox{0.34}[0.32]{\includegraphics{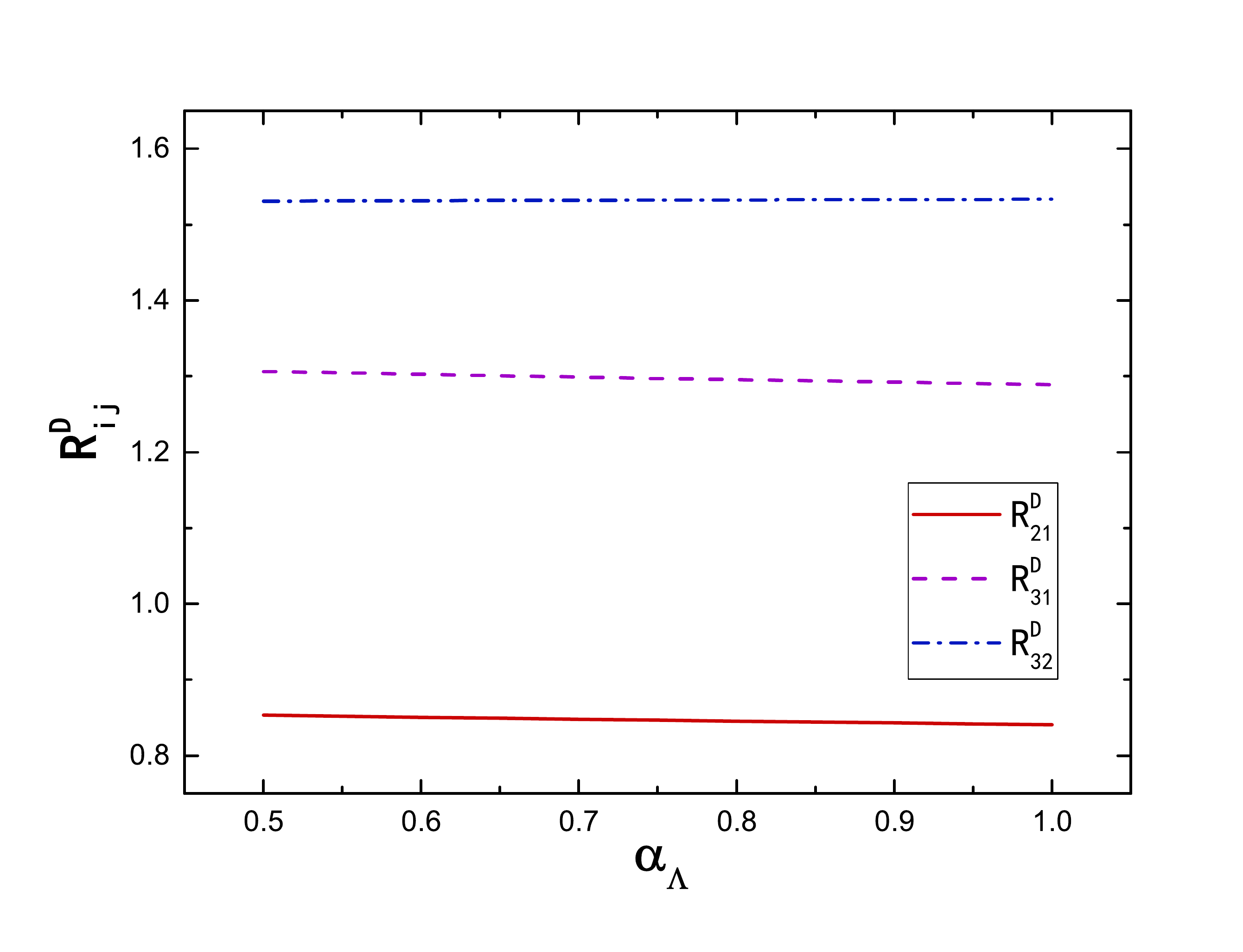}}
\caption{The $\alpha_\Lambda$ dependence of the ratios $\mathcal{R}^D_{21} = \mathcal{B}[\Upsilon(6S) \to \Upsilon(1^3D_2) \eta] / \mathcal{B}[\Upsilon(6S) \to \Upsilon(1^3D_1) \eta$, $\mathcal{R}^D_{31} = \mathcal{B}[\Upsilon(6S) \to \Upsilon(1^3D_3) \eta] / \mathcal{B}[\Upsilon(6S) \to \Upsilon(1^3D_1) \eta]$, and $\mathcal{R}^D_{32} = \mathcal{B}[\Upsilon(6S) \to \Upsilon(1^3D_3) \eta] / \mathcal{B}[\Upsilon(6S) \to \Upsilon(1^3D_2) \eta]$.}
\label{fig:B321-13DJ}
\end{figure}
\end{center}

With above preparation, we now evaluate the branching ratios of the $\Upsilon(6S) \to \Upsilon(1^3D_J) \eta$ transitions.
In Ref. \cite{Wang:2016qmz}, we have predicted the branching ratios of the $\Upsilon(5S)\to \Upsilon(1^3D_J)\eta$ by considering the hadronic loop mechanism, which are consistent with the experimental measurement given by Belle later \cite{Tamponi:2018cuf}. In this theoretical calculation \cite{Wang:2016qmz}, $\alpha_\Lambda=0.5\sim 1.0$ was taken. Due to the similarity between $\Upsilon(5S)\to \Upsilon(1^3D_J)\eta$ and $\Upsilon(6S)\to \Upsilon(1^3D_J)\eta$, in this work we set the same $\alpha_\Lambda$ range to predict the decay behaviors of $\Upsilon(6S)\to \Upsilon(1^3D_J)\eta$.
In  Fig.~\ref{fig:B123-13DJ}, we illustrate the $\alpha_\Lambda$ dependence of the branching ratios,
while in Fig.~\ref{fig:B321-13DJ}, the $\alpha_\Lambda$ dependence of the relative ratios among these branching fractions is presented.

Our calculation shows that the branching ratios of $\Upsilon(6S) \to \Upsilon(1^3D_J) \eta$ shown in Fig. \ref{fig:B123-13DJ} can reach up to $10^{-3}$, i.e.,
\begin{eqnarray}\label{1D}
\mathcal{B}[\Upsilon(6S) \to \Upsilon(1^3D_1) \eta] &=& (0.174 \sim 1.720) \times 10^{-3} , \nonumber\\
\mathcal{B}[\Upsilon(6S) \to \Upsilon(1^3D_2) \eta] &=& (0.149 \sim 1.440) \times 10^{-3} , \nonumber\\
\mathcal{B}[\Upsilon(6S) \to \Upsilon(1^3D_3) \eta] &=& (0.228 \sim 2.210) \times 10^{-3}.\nonumber
\end{eqnarray}
These significant values of the branching ratios show that the $\Upsilon(6S) \to \Upsilon(1^3D_J) \eta$ decays can be accessible at Belle II.

We also obtain three ratios as shown in Figs. \ref{fig:B321-13DJ}, where
these ratios weakly depend on $\alpha_\Lambda$. The predicted ratios include
\begin{eqnarray}\label{ratios}
\mathcal{R}^D_{21} &=& \frac{\mathcal{B}[\Upsilon(6S) \to \Upsilon(1^3D_2) \eta]}{\mathcal{B}[\Upsilon(6S) \to \Upsilon(1^3D_1) \eta]} \approx 0.841 \sim 0.853, \nonumber\\
\mathcal{R}^D_{31} &=& \frac{\mathcal{B}[\Upsilon(6S) \to \Upsilon(1^3D_3) \eta]}{\mathcal{B}[\Upsilon(6S) \to \Upsilon(1^3D_1) \eta]} \approx 1.289 \sim 1.306, \nonumber\\
\mathcal{R}^D_{32} &=& \frac{\mathcal{B}[\Upsilon(6S) \to \Upsilon(1^3D_3) \eta]}{\mathcal{B}[\Upsilon(6S) \to \Upsilon(1^3D_2) \eta]} \approx 1.531 \sim 1.533, \nonumber
\end{eqnarray}
which can be tested further in future experiment like Belle and Belle II.

\section{Summary}

Since 2007, Belle has paid more attentions to the hidden-bottom hadronic transitions of the $\Upsilon(5S)$, and has found
anomalous hadronic transitions like $\Upsilon(5S)\to \Upsilon(nS)\pi^+\pi^-$ ($n=1,2,3$) \cite{Abe:2007tk}
and $\Upsilon(5S)\to \chi_{bJ}\omega$ ($J=0,1,2$) \cite{He:2014sqj}. Targeted on these anomalies, theorists have made great efforts to explore the underlying mechanisms \cite{Chen:2011zv,Meng:2007tk,Meng:2008dd,Simonov:2008qy,Chen:2011qx,Chen:2014ccr,Chen:2011pv,Meng:2008bq,Wang:2016qmz}, and have revealed that the hadronic loop mechanism, which is an equivalent description of the coupled-channel effect, plays an important role for understanding these anomalous transitions of $\Upsilon(5S)$.

Besides these observations, Belle discovered the $\Upsilon(5S)$ decays into $\Upsilon(1^3D_J)\eta$ very recently, which 
has large branching ratios \cite{Tamponi:2018cuf}. In fact, this Belle measurement confirmed the prediction made in Ref. \cite{Wang:2016qmz}, which again shows the important role of the hadronic loop effect to the $\Upsilon(5S)$ hidden-bottom decays. What is more important is that this fact also inspires our ambition to further explore the hidden-bottom decays of the $\Upsilon(6S)$.

In this work, we have selected the $\Upsilon(6S)\to \Upsilon(1^3D_J)\eta$, and have studied the potential discovery of these decays in a future experiment. Our calculation has shown that the branching ratios of the $\Upsilon(6S)\to \Upsilon(1^3D_J)\eta$ can reach up to $10^{-3}$ when the hadronic loop effect is introduced. It is evident that these significant branching ratios could arouse experimentalist's interest in finding them.

With the running of Belle II, we expect the observation of these anomalous $\Upsilon(6S)\to \Upsilon(1^3D_J)\eta$ transitions, which will make our understanding of higher bottomonia more in-depth and thorough. Notably, the present study should become a part of the whole research around the $\Upsilon(6S)$.

\section*{Acknowledgments}

Qi Huang would like to thank Bo Wang for useful discussions. This project is supported by the National Natural Science Foundation of China under Grants No.~11222547 and No.~11175073.
Xiang Liu is also supported by the National Program for Support of Young Top-notch Professionals.

\vfil

\end{document}